\documentclass[
    ,final            
  ]
  {aipproc}

\layoutstyle{6x9}

\usepackage{amsmath,amssymb}

\begin{document}

\title{\vspace*{-2cm} \hspace{10cm}{\small \hfill{DCPT/10/110}}\\[-0.3cm]\hspace{10.35cm}{\small \hfill{IPPP/10/55}}\\[1cm]Notes from the 3rd Axion Strategy Meeting}

\classification{14.80.-j, 14.80.Va} \keywords      {Strategy for future axion
and general low energy particle physics}

\author{O.K. Baker}{
  address={Department of Physics, Yale University, P.O. Box 208120, New Haven, CT 06520}
}

\author{G. Cantatore}{
  address={Universitá and INFN Trieste, via Valerio 2, 34127 Trieste, Italy }
}

\author{J. Jaeckel}{
  address={Institute for Particle Physics Phenomenology, Durham University, Durham DH1 3LE, UK}
}

\author{G. Mueller}{
  address={Department of Physics, University of Florida, PO Box 118440, Gainesville, FL 32611, USA}
}

\begin{abstract}
In this note we briefly summarize the main future targets and strategies
for axion and general low energy particle physics identified in the
\textquotedblleft{}3rd axion strategy meeting\textquotedblright{}
held during the AXIONS 2010 workshop. This summary follows a wide
discussion with contributions from many of the workshop attendees.
\end{abstract}

\maketitle

\section{Introduction}

Low energy experiments can be a powerful probe of particle physics
beyond the current Standard Model. The classic examples are experiments
searching for axions such as cavity searches for dark matter axions,
axion helioscopes, light shining through a wall experiments as well as laser polarization experiments. Over the last few years it has
also been realized that these and similar experiments can indeed provide
an interesting probe not only of axions, but of a wide variety of
particles suggested in realistic extensions of the Standard Model.
In addition to the strong-as-ever motivation to find the axion this
has significantly strengthened the physics case for these experiments
and has also led to a general stepping up of all the experimental
efforts.

The purpose of the ``3rd axion strategy meeting'' was to provide a
round table environment for the community of physicists, both theorists
and experimentalists, working in axion, and more generally in WISP
physics, to freely float ideas and informally discuss them in order
to discern the global trend of the field and identify interesting
and promising directions for future developments. In this note we
will briefly summarize the results of the discussion.

The ADMX experiment is currently the only experiment sensitive enough to probe
the preferred region for the QCD axion (which is not yet excluded by astrophysics) and in particular also one of the
regions suggested by astrophysical puzzles
-- the region where axion is dark matter. But the ADMX experiment
is intrinsically narrowband and can not probe the entire mass range
in any reasonable time \cite{Rosenberg,Jesse}.
Moreover, ADMX is based on the assumption that a significant fraction of the dark matter is made of axions -- an assumption
that is natural only in a limited range of axion parameters.
The discussion focussed on the status of the
current experiments \cite{CAST, Tokyo, ALPS}, potential
ways to improve the sensitivity of the optical experiments \cite{Gui, GuiSik}
to complement and go beyond the ADMX experiment
in particular by probing larger mass ranges. Currently, most experiments are rather small
scale and pursued by relatively small groups with little to no hope
to reach the QCD range in the near future. However, they develop the
technologies and gain the experience needed to later develop and
build a scaled up experiment which might potentially reach the QCD
range. The question was also raised if this requires already a
global initiative merging to build a single experiment.

\section{Heading towards the axion line}

Although important new motivation in the form of a variety of hidden
sector particles has arisen over the last few years, one of the main
targets of the community remains to discover the standard QCD axion.
The following discussion therefore focussed on the potential improvement of axion searches. However, it should
be stressed that basically all improvements in axion searches directly translate into similar improvements for general WISP searches.
Furthermore, it should also be noted that for some WISPs such as, e.g. hidden sector photons the laboratory experiments
already probe interesting parameter space untested by astrophysics. Therefore, experiments working towards reaching the axion
line receive very good motivation from their significant discovery potential for these particles.

\begin{figure}
  \includegraphics[height=.3\textheight]{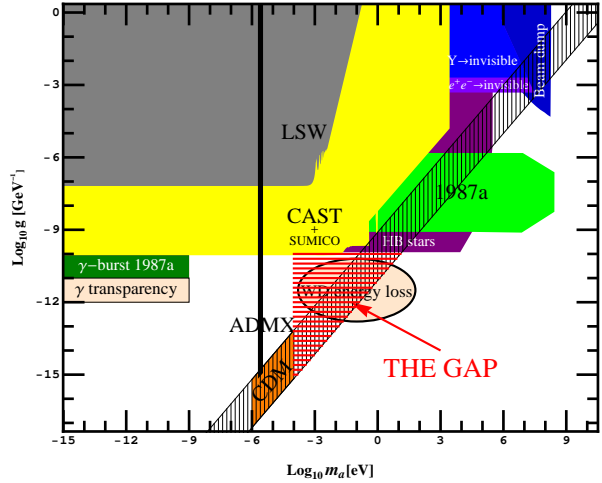}
  \caption{Current experimental and astrophysical constraints on axion-like particles (cf., e.g., Ref.\cite{Joerg, Jaeckel:2010} for details). (Light) orange areas denote astrophysical hints.
  The hatched area is currently unexplored and it is the target area for future laboratory searches for the QCD axion.}
  \label{axionfig}
\end{figure}

Strong astrophysical bounds require that the future laboratory
experiments that can successfully probe the QCD axion must significantly
improve in sensitivity towards smaller couplings but also towards
larger masses (see Fig.\ref{axionfig}). The main
technologies discussed to close the gap were light-shining-through-walls
(LSW) experiments and helioscopes.

Helioscopes \cite{CAST, Tokyo} use a strong magnetic field to convert relativistic
solar axions to X-rays. CAST \cite{CAST} has reached a sensitivity in coupling of
$10^{-10}$~GeV$^{-1}$ for masses smaller than $20$~meV
and is currently the most sensitive broadband observatory-type experiment in the world.
Improvements could come from advances in
the efficiency of the X-ray telescopes and from reductions in the dark
current of the X-ray detectors. However, fundamentally, the sensitivity
scales with the magnetic field, the length, and the cross sectional
area. Increases in the length and cross sectional area will reach
practical limits as the telescope has to track the sun to detect solar
axions or will only be sensitive during short periods of time. The
resonant amplification of the generated X-rays was also disregarded
as the solar axions are incoherent and any resonant amplification
in one frequency region will suppress signals in other frequency regions.
The net effect would be a reduction in the signal.  The main avenue of approach here leads to stronger magnetic 
fields and perhaps wider apertures.

LSW experiments first convert an optical photon into a relativistic
axion inside a strong magnetic field. The axion then propagates through
'a wall' into a second strong magnetic field region where it converts
back into an optical photon. Over the last years LSW experiments have
improved considerably. The next step seems to be the use of optical
cavities in the production and in the regeneration regions. In the
production region this has been pioneered by the ALPS experiment\footnote{It
should be noted that BFRT also had mirrors to enhance the production
probability, but the setup used was an optical delay line and not
a cavity}. Installing a second resonant cavity in the regeneration
region seems to be a viable option \cite{Hoogeveen:1990vq,Sikivie:2007} and many
groups are pursuing it. An example is the effort by a collaboration
led by groups at Fermilab and the University of Florida to set up
a resonant regeneration experiment \cite{Gui}. The goal is using 6+6 Tevatron
magnets (5 T field, 48 mm bore diameter). The present stage is proposal
preparation and a pilot version of the setup with 2 m + 2 m long 0.5
T permanent magnets which is forseen for the near future at the University
of Florida. This LSW experiment can in principle reach sensitivities
in the $10^{-11}\mbox{GeV}^{-1}$ range and will ultimately be limited
by the ability to handle the laser power and optical losses inside
the cavities.

Laser polarization (LPol) experiments can in principle reach sensitivities
similar to light-shining-through-walls (LSW) experiments.
In these experiments the detection of the axion coming from the conversion of a linearly polarized photon in an external field is based on sensing either the delay experienced by the photon when it briefly oscillates into a massive axion, or the disappearance of the photon itself due to the production of a real axion. The delay causes birefringence, while the disappearance causes an apparent rotation, actually a dichroism.

LPol experiments have the advantage that only
one long interaction region is required. However, the birefringence
caused by QED effects in vacuum creates a background which would be
larger and indistinguishable from an axion signal, unless birefringence measurements are combined with rotation measurements. Also, in practice, all the necessary optical elements have an intrinsic birefringence which severely limits the attainable sensitivity.
For these reasons LPol experiments appear to be less promising than LSW experiments. 

In cavity experiments like ADMX using cold dark matter axions, the mass and the resonance frequency of the cavity have to agree within the bandwidth of the cavity in order to achieve an enhancement of the signal due to resonant regeneration. For a given cavity frequency the experiment is sensitive only in a very limited mass bandwidth. In order to explore an appreciable range of the, a priori unknown, axion mass, one has to measure cavity power after tuning the cavity at a given frequency, then slightly change the frequency of the cavity, measure again and so on in order to scan over the axion masses. This process takes time. With the next stage of ADMX, roughly a decade of axion masses will be explored, and future developments may allow to extend this by another decade. Extension beyond the mass range $1-100\,\mu$eV seems challenging.
Moreover, cavity searches for dark matter axions build on a strong assumption, namely that  all (or at least a sizeable part) of the dark matter is axions.
In contrast, LSW or helioscope experiments have their own production mechanisms and do not rely on this assumption.

Following these arguments, LSW and helioscope experiments appear to
be the most promising experiments beyond ADMX. Both of them require
stronger magnets and larger interaction regions to reach the QCD axion
range. But there are differences in the way the two experiments scale
with size. While the sensitivity of the helioscope experiments scale
with the cross sectional area, LSW experiments only require cross
sections which exceed the cross sections of the diverging laser beams.
Scaling up the length of both types of experiments would improve the
sensitivity but at the same time reduce the probed mass range. This
mass range can be shifted towards higher masses using techniques to
maintain the phase matching between the electrical and the axion field.
These techniques could include refractive materials inside the cavity
such as a dilute gas or phase shifting glass plates or by alternating
(periodically poling) the magnetic field. Note that the introduction
of material into the LSW cavities will increase the optical losses
and will limit the finesse of the cavities.

In addition to the above, new frequency regimes may be explored. For
example helioscopes could also search in the eV regime. Currently,
calculations of the flux are underway. Moreover, there are also efforts
to run an LSW experiment with microwave cavities. This promises good
sensitivity towards small couplings. However, the mass range is somewhat
limited by the lower frequency/energy radiation.

\section{A global design initiative}

An additional area of intense discussion was the future ``strategy''
of the community. The development of stronger and larger magnets appears
to be the key to increase the sensitivity of helioscope and LSW experiments
into the QCD axion range. However, the longer and stronger approach
faces two main challenges. First of all the development of stronger
magnets is costly both in time and in money. Yet, here the axion searches
may profit from developments from other programs such as the \textquotedbl{}new
intensity frontier\textquotedbl{} program at Fermilab, where magnets
with field intensities up to $14$~T are studied. Our
community should approach Fermilab and other labs (such as CERN) involved in magnet development  and request that these magnets
have large enough sufficient apertures for our experiments. This request
will carry more weight if our community acts as a coherent community
and promotes only a very limited number of potential experimental
designs. This is not the case at present.

Secondly, as discussed above longer magnets increase the sensitivity
towards smaller couplings for low mass axions but decrease the sensitivity
for larger masses making it even more difficult to reach the QCD axion
line. The development of potential phase matching techniques or alternative
designs appears to be necessary before any large scale experiments
beyond the currently planned experiments can proceed.

Furthermore, increasing size (and cost) of the experiments does require
a consolidation of the experimental efforts, preceded by the convergence
on a main production/detection technique. The extreme form would be
the selection of a single site for a large scale experiment. Yet,
at the same time a diverse community of experimental groups is probably
a very positive feature to explore a variety of experimental techniques.
The general agreement at the end was that the \textquotedbl{}world
experiment\textquotedbl{} should for the moment be left as an open
issue. It would also be useful to prepare a summary document with
ideas and relevant numbers which could be used for funding agencies.

\bibliographystyle{aipproc}   

\end{document}